\documentclass[twocolumn,tighten]{aastex701} %linenumbers,

\usepackage{amsmath}
\usepackage{graphicx}

\usepackage[symbol]{footmisc}
\usepackage{multirow}
\usepackage{booktabs}
\usepackage{xcolor}
\usepackage{soul}

\def\arraystretch{1.19}

\RequirePackage{color}
\definecolor{cblue}{cmyk}{55,17,0,0}
\definecolor{darkgreen}{rgb}{0.0, 0.4, 0.0}
\definecolor{darkorange}{rgb}{0.5, 0.3, 0.0}
\definecolor{darkblue2}{rgb}{0.17, 0.49, 0.72}

\newcommand{\sunrise}{\mbox{{\sc Sunrise~iii}}}

\newcommand{\hdelta}{H$\delta$}

\newcommand{\weaklines}{46}

\usepackage{caption}
\DeclareCaptionFormat{cont}{#1 (cont.)#2#3\par}

\newcommand{\animation}{
https://www.dropbox.com/scl/fi/olokk1vmvhjpfms5iy286/Sunrise_limb_obs_16Jul24_12UT.mp4?rlkey=4y7i8ag57bq41ez7uxwgdixka&st=iq8mcqx4&dl=0}

\newcommand{\WavelengthRange}{407.60--410.24}

\newcommand{\TotalElements}{41}
\newcommand{\TotalLines}{199}

\newcommand{\MaxAtomicNumber}{72}

\newcommand{\TotalLinesOndiskOnly}{25}
\newcommand{\TotalLinesOffLimbOnly}{34}
\newcommand{\TotalLinesOnBoth}{140}
\newcommand{\TotalLinesOffdiskOnly}{174}

 \newcommand{\PctIonizedLines}{63.8}
 \newcommand{\PctNeutralLines}{36.2}

\newcommand{\UnidentifiedLinesCount}{34}
\newcommand{\PctUnidentifiedLines}{17.6\,\%}

\newcommand{\totalLinesEmissionOndisk}{122}
\newcommand{\totalLinesEmissionOndiskIdentified}{98}
\newcommand{\PcttotalLinesEmissionOndisk}{61.3\,\%}

\newcommand{\PtctotalLinesEmissionOndiskIdentifiedionized}{65.3\,\%}

\newcommand{\totalLinesEmissionOndiskTBI}{24}

\graphicspath{{figures/}}

\begin{document}
\correspondingauthor{Juan~Sebastián~Castellanos~Durán}

\author[orcid=0000-0003-4319-2009,sname='Castellanos~Durán']{Juan~Sebastián~Castellanos~Durán}
\affiliation{Max-Planck-Institut für Sonnensystemforschung, Justus-von-Liebig-Weg 3, 37077 Göttingen, Germany}
\email[show]{castellanos@mps.mpg.de}

\author[orcid=0000-0003-1409-1145,sname='Iglesias']{Francisco~A.~Iglesias} \affiliation{Max-Planck-Institut für Sonnensystemforschung, Justus-von-Liebig-Weg 3, 37077 Göttingen, Germany}\affiliation{Grupo de Estudios en Heliofísica de Mendoza, CONICET, Universidad de Mendoza, Boulogne sur Mer 683, 5500 Mendoza, Argentina}\email{iglesias@mps.mpg.de}

\author[orcid=0000-0002-3418-8449,sname='Solanki']{Sami~K.~Solanki} \affiliation{Max-Planck-Institut für Sonnensystemforschung, Justus-von-Liebig-Weg 3, 37077 Göttingen, Germany}\email{solanki@mps.mpg.de}	

\author[orcid=0000-0003-1459-7074,sname='Lagg']{Andreas~Lagg} \affiliation{Max-Planck-Institut für Sonnensystemforschung, Justus-von-Liebig-Weg 3, 37077 Göttingen, Germany}\email{lagg@mps.mpg.de}	

\author[orcid=0000-0002-0189-5550,sname='Milic']{Ivan~Milic}
\affiliation{Institut für Sonnenphysik (KIS), Georges-Köhler-Allee 401a, 79110 Freiburg, Germany}
\email{milic@leibniz-kis.de}

\author[orcid=0000-0003-3490-6532,sname='Smitha']{H.~N.~Smitha}
\affiliation{Max-Planck-Institut für Sonnensystemforschung, Justus-von-Liebig-Weg 3, 37077 Göttingen, Germany}
\email{narayanamurthy@mps.mpg.de}

\author[orcid=0000-0003-1971-5551,sname='Vukadinović']{Dušan~Vukadinović}
\affiliation{Max-Planck-Institut für Sonnensystemforschung, Justus-von-Liebig-Weg 3, 37077 Göttingen, Germany}
\email{vukadinovic@mps.mpg.de}

\author[orcid=0000-0001-6317-4380,sname='Riethmüller']{Tino~L.~Riethmüller}
\affiliation{Max-Planck-Institut für Sonnensystemforschung, Justus-von-Liebig-Weg 3, 37077 Göttingen, Germany}
\email{riethmueller@mps.mpg.de}

\author[orcid=0000-0002-9972-9840,sname='Gandorfer']{Achim~Gandorfer} \affiliation{Max-Planck-Institut für Sonnensystemforschung, Justus-von-Liebig-Weg 3, 37077 Göttingen, Germany}\email{gandorfer@mps.mpg.de}

\author[orcid=0009-0009-4425-599X,sname='Feller']{Alex~Feller} \affiliation{Max-Planck-Institut für Sonnensystemforschung, Justus-von-Liebig-Weg 3, 37077 Göttingen, Germany}\email{feller@mps.mpg.de}

\author[orcid=0000-0002-2055-441X,sname='Blanco~Rodríguez']{Julian~Blanco~Rodríguez}
\affiliation{Universitat de Valencia Catedrático José Beltrán 2, E-46980 Paterna-Valencia, Spain}
\affiliation{Spanish Space Solar Physics Consortium}
\email{julian.blanco@uv.es}

\author[orcid=0000-0001-8829-1938,sname='Orozco~Suárez']{David~Orozco~Suárez}
\affiliation{Instituto de Astrofísica de Andalucía, CSIC, Glorieta de la Astronomía s/n, 18008 Granada, Spain}
\affiliation{Spanish Space Solar Physics Consortium}
\email{orozco@iaa.es}

\author[orcid=0000-0002-5054-8782,sname='Katsukawa']{Yukio~Katsukawa}
\affiliation{National Astronomical Observatory of Japan, 2-21-1 Osawa, Mitaka, Tokyo 181-8588, Japan}
\affiliation{Department of Earth and Planetary Science, The University of Tokyo, 7-3-1, Hongo, Bunkyo-ku, Tokyo 113-0033, Japan}
\affiliation{Department of Astronomical Science, The Graduate University for Advanced Studies (SOKENDAI), 2-21-1 Osawa, Mitaka, Tokyo 1818588, Japan}
\email{yukio.katsukawa@nao.ac.jp}

\author[orcid=0000-0001-6029-7529,sname='Hoelken']{Johannes~H\"olken}
\affiliation{Max-Planck-Institut für Sonnensystemforschung, Justus-von-Liebig-Weg 3, 37077 Göttingen, Germany}
\email{hoelken@mps.mpg.de}

\author[orcid=0000-0002-3387-026X,sname='del~Toro~Iniesta']{Jose~Carlos~del~Toro~Iniesta}
\affiliation{Instituto de Astrofísica de Andalucía, CSIC, Glorieta de la Astronomía s/n, 18008 Granada, Spain}
\affiliation{Spanish Space Solar Physics Consortium}
\email{jti@iaa.es}

\author[orcid=0000-0002-7544-6926,sname='Rao']{Shihao~Rao}
\affiliation{Max-Planck-Institut für Sonnensystemforschung, Justus-von-Liebig-Weg 3, 37077 Göttingen, Germany}
\email{sh_rao@163.com}

\author[orcid=0009-0008-6857-9043,sname='Sanchez~Toledo']{Mariano~Sanchez~Toledo}
\affiliation{Grupo de Estudios en Heliofísica de Mendoza, CONICET, Universidad de Mendoza, Boulogne sur Mer 683, 5500 Mendoza, Argentina}
\email{mt.sanchez@alumno.um.edu.ar}

\author[orcid=0009-0007-5514-4553,sname='Singh']{Kunal~Singh}
\affiliation{Max-Planck-Institut für Sonnensystemforschung, Justus-von-Liebig-Weg 3, 37077 Göttingen, Germany}
\email{singhk@mps.mpg.de}

\author[orcid=0000-0002-0787-8954,sname='Bernasconi']{Pietro~Bernasconi}
\affiliation{Johns Hopkins University Applied Physics Laboratory, 11100 Johns Hopkins Road, Laurel, Maryland, USA}
\email{pietro.bernasconi@jhuapl.edu}

\author[sname='Berkefeld']{Thomas~Berkefeld}
\affiliation{Institut für Sonnenphysik (KIS), Georges-Köhler-Allee 401a, 79110 Freiburg, Germany}
\email{thomas.berkefeld@leibniz-kis.de}

\author[orcid=0000-0001-9228-3412,sname='Álvarez-Herrero']{Alberto~Álvarez-Herrero}
\affiliation{Instituto Nacional de T\'ecnica Aeroespacial (INTA), Ctra. de Ajalvir, km. 4, E-28850 Torrejón de Ardoz, Spain}
\affiliation{Spanish Space Solar Physics Consortium}
\email{alvareza@inta.es}

\author[orcid=0000-0001-5616-2808,sname='Kubo']{Masahito~Kubo}
\affiliation{National Astronomical Observatory of Japan, 2-21-1 Osawa, Mitaka, Tokyo 181-8588, Japan}
\email{masahito.kubo@nao.ac.jp}

\author[orcid=0000-0001-7764-6895,sname='Martínez~Pillet']{Valentín~Martínez~Pillet}
\affiliation{Instituto de Astrofísica de Canarias, Vía Láctea, s/n, E-38205 La Laguna, Spain}
\affiliation{Spanish Space Solar Physics Consortium}
\email{vmpillet@iac.es}

\author[sname='Grauf']{Bianca~Grauf}
\affiliation{Max-Planck-Institut für Sonnensystemforschung, Justus-von-Liebig-Weg 3, 37077 Göttingen, Germany}
\email{grauf@mps.mpg.de}

\author[sname='Carpenter']{Michael~Carpenter}
\affiliation{Johns Hopkins University Applied Physics Laboratory, 11100 Johns Hopkins Road, Laurel, Maryland, USA}
\email{michael.carpenter@jhuapl.edu}

\author[sname='Bell']{Alexander~Bell}
\affiliation{Institut für Sonnenphysik (KIS), Georges-Köhler-Allee 401a, 79110 Freiburg, Germany}
\email{albe@leibniz-kis.de}

\author[orcid=0000-0003-1483-4535,sname='Strecker']{Hanna~Strecker}
\affiliation{Instituto de Astrofísica de Andalucía, CSIC, Glorieta de la Astronomía s/n, 18008 Granada, Spain}
\affiliation{Spanish Space Solar Physics Consortium}
\email{streckerh@iaa.es}

\author[orcid=0000-0003-0175-6232,sname='Siu-Tapia']{Azaymi~L.~Siu-Tapia}
\affiliation{Instituto de Astrofísica de Andalucía, CSIC, Glorieta de la Astronomía s/n, 18008 Granada, Spain}
\affiliation{Spanish Space Solar Physics Consortium}
\email{siu@iaa.es}

\author[orcid=0000-0001-7094-518X,sname='Santamarina~Guerrero']{Pablo~Santamarina~Guerrero}
\affiliation{Instituto de Astrofísica de Andalucía, CSIC, Glorieta de la Astronomía s/n, 18008 Granada, Spain}
\affiliation{Spanish Space Solar Physics Consortium}
\email{psanta@iaa.es}

\author[orcid=0000-0001-7452-0656,sname='Kawabata']{Yusuke~Kawabata}
\affiliation{National Astronomical Observatory of Japan, 2-21-1 Osawa, Mitaka, Tokyo 181-8588, Japan}
\email{kawabata.yusuke@nao.ac.jp}

\author[orcid=0000-0002-1043-9944,sname='Matsumoto']{Takuma~Matsumoto}
\affiliation{Centre for Integrated Data Science, Institute for Space-Earth Environmental Research, Nagoya University, Furocho, Chikusa-ku, Nagoya, Aichi 464-8601, Japan}
\email{takuma.matsumoto@gmail.com}

\author[orcid=0000-0002-4669-5376,sname='Ishikawa']{Ryohtaroh~T.~Ishikawa}
\affiliation{National Institute for Fusion Science, 322-6 Oroshi-cho, Toki City 509-5292, Japan}
\email{ishikawa.ryohtaro@nifs.ac.jp}

\author[orcid=0009-0002-6808-5154,sname='Harnes']{Edvarda~Harnes}
\affiliation{Max-Planck-Institut für Sonnensystemforschung, Justus-von-Liebig-Weg 3, 37077 Göttingen, Germany}
\email{harnes@mps.mpg.de}

\author[orcid=0000-0002-7044-6281,sname='Oba']{Takayoshi~Oba}
\affiliation{Max-Planck-Institut für Sonnensystemforschung, Justus-von-Liebig-Weg 3, 37077 Göttingen, Germany}
\affiliation{National Astronomical Observatory of Japan, 2-21-1 Osawa, Mitaka, Tokyo 181-8588, Japan}
\email{oba@mps.mpg.de}

% \author[]{{\textbf{et al.}}}
% \affiliation{}
% \email{}

% \collaboration{32}{{\sunrise{} consortium}}

\suppressAffiliations

\title{A forest of emission lines observed by Sunrise III/SUSI above the solar limb}

\received{26/08/2026}
\revised{17/09/2026} 
\accepted{19/09/2026}

\begin{abstract}
We present off-limb spectroscopy using low-stray-light data from the Sunrise Spectropolarimeter and Imager (SUSI) on board the \sunrise{} stratospheric solar observatory. A spatial scan with SUSI's slit almost parallel to the local limb reached from within the solar disk to 6.5 Mm off-limb.  The wavelength range spanned from 407.6 to 410.25\,nm. We detected a total of \TotalLinesOffdiskOnly{} spectral lines in emission off-limb, a 2.5-fold increase over previous reports. Surprisingly, \totalLinesEmissionOndisk{} of these lines begin to appear in emission already on the disk, although close to the limb. In addition, one-fifth of relatively prominent emission lines are due to rare-earth and heavy elements and molecules, where the absorption lines of these transitions are typically weak on-disk.
\end{abstract}


\section{Introduction}\label{sec:intro}

Spectroscopy in visible wavelengths close to and beyond the solar limb opens a unique window onto the solar atmosphere, allowing spectra to be probed in a very different geometry than on the solar disk. Low contrast resulting from diffuse sky radiance due to atmospheric Rayleigh scattering, along with variable seeing conditions, makes such spectroscopic observations challenging from the ground \citep{Penn1996SoPh..Limb...He10830}. Different approaches have been taken to circumvent these problems. 

Solar eclipses offer a natural setting for observations with reduced on-disk stray light. Flash spectra, visible for only a few seconds during the beginning and the end of totality, reveal off-limb emission lines \citep[e.g.,][and references therein]{Judge2019SoPh..Eclipse}.
Outside solar eclipses, good-quality spectroscopic observations covering photospheric lines are extremely rare and often lack detailed spatial resolution along the radial direction across the limb \citep[][ and references therein]{Pierce1968ApJS...ChromosphericSpectrum, Shen1981ChA&A...Limb...Flashspectrum}. 

An alternative to ground-based measurements is to perform space-based observations. The spectropolarimeter \citep{Ichimoto2008SoPh, Tsuneta2008} onboard Hinode \citep{Kosugi2007} observes in the photospheric iron-line pair around 630\,nm and also at the extreme limb and off-limb. As the line of sight approaches the extreme solar limb, the steep drop in continuum opacity alters spectral-line formation, driving lines from absorption into emission. This phenomenon is well known for strong chromospheric lines \citep[e.g.,][]{Livingston2011SoPh..WeakEmissionLine, Beck2011A&A...Limb,Tei2020ApJ...Off-limbSpicules,Tei2025...LimbCoronalholequietIRIS}. Notably, \citet{Lites2010ApJ...Limb} demonstrated that the photospheric Fe I 630\,nm pair goes into emission just above the limb. 

Even with recent upgrades to adaptive optics systems that allow locking near or off the limb \citep[e.g.,][]{Schmidt2025NatAs...AO.GST.Halpha, Kuridze2019ApJ...flarelimb}, high-quality limb spectra combining high spectral and spatial resolution with sufficiently low stray light remain rare when obtained without an occulter. While near- and off-limb observations in general are not scarce, such noncoronagraphic data are difficult to acquire and mainly restricted to chromospheric lines. This must be distinguished from low-stray-light coronagraphic spectroscopy where the solar disk is occulted \citep[e.g.,][]{Morton2025ApJ...DKIST}, which has enabled, for example, observations of forbidden infrared coronal lines \citep{Schad2023ApJ...DKISTCorona}. The very low flux reaching the detector in the ultraviolet, due to optical throughput and absorption and Rayleigh scattering  in Earth's atmosphere, which depends on $I_{\rm Rayleigh}\propto 1/\ \lambda^{4}$, further complicates matters for measurements in the violet.

\begin{figure*}
    \centering
    \includegraphics[width=0.99\textwidth]{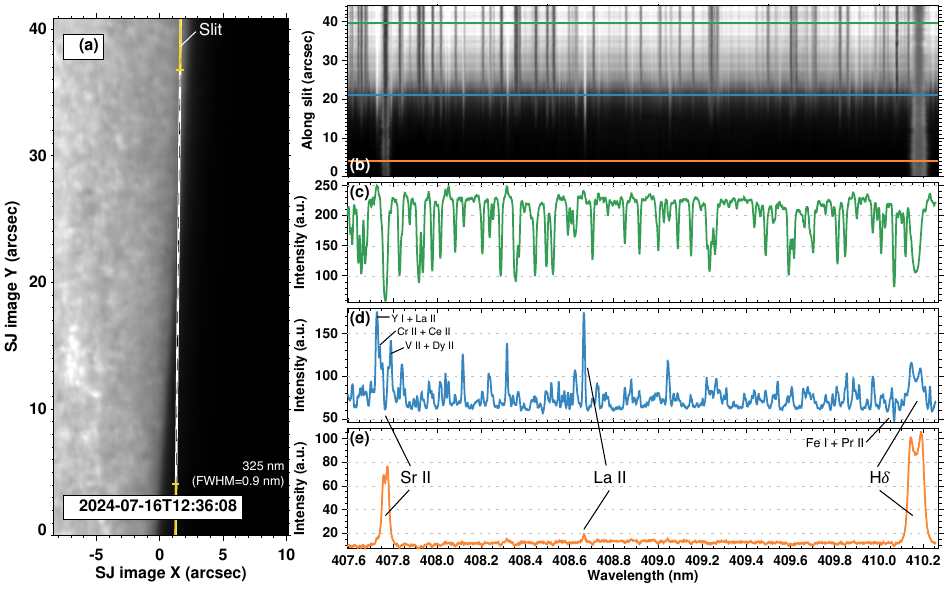}
    \caption{SUSI data recorded around the solar limb. Panel (a) shows a portion of SUSI's broadband slit-jaw (SJ) channel at 325\,nm and a full width at half-maximum (FWHM) of 0.9\,nm \citep{Feller2025SoPh..SUSI}. The white and yellow marks show the location of the slit at the time inscribed on the panel. Panel (b) is the spectrum, and panels (c-e) are cuts along the green, blue, and orange lines marked in panel (b). An \href{\animation{}}{animation} of the whole observation is part of the online material.}\label{fig:observations}
\end{figure*}

Another approach has been to observe in the infrared \citep[e.g.,][]{Solanki1994Sci...DarknessChromosphere}. Observations at the Vacuum Tower Telescope (VTT) identified off-limb signatures of spicules in strong chromospheric lines such as the \ion{Ca}{2} infrared line at 854.2\,nm or in the \ion{He}{1} triplet at 1083.0, the former vanishing at about 5\,Mm and the latter extending up to 6\,Mm above the limb \citep{Beck2016SoPh...Limb}.

The \textsc{Sunrise} observatory overcame these challenges by having a 1-meter telescope flying on a stratospheric balloon to obtain observations in a seeing-free environment at an altitude of 35 km above ground \citep{Solanki2010ApJ...Sunrise, Solanki2017ApJS...Sunrise, Barthol2011SoPh...Sunrise,Korpi-Lagg2025...SunriseIII}. 

In this Letter, we analyze observations taken by \sunrise{} during its third science flight in 2024 \citep{Solanki2026...sunrise}. Thanks to the high-precision pointing \citep{Bernasconi2025SoPh...Sunrise} and image stabilization system \citep[CWS;][]{Berkefeld2026...sunrise}, \sunrise{} conducted continuous, hour-long spectropolarimetric observations from the ultraviolet to the infrared \citep{Feller2025SoPh..SUSI, delToroIniesta2025SoPh...TuMag, Katsukawa2026...sunrise} at the solar limb and beyond \citep[see e.g.,][]{Naito2026sunrise, Haocheng2026ApJL, Milic2026sunrise}.

We present spectroscopic analysis of observations acquired with the \textsc{Sunrise} UV Spectropolarimeter and Imager \citep[SUSI;
][]{Feller2025SoPh..SUSI, Iglesias2025SoPh...Sunrise} and report a forest of emission lines observed at the extreme-limb and off-limb regions, the majority of which are photospheric. We study the height variation of the intensity in hundreds of spectral lines around 408\,nm, including hydrogen Balmer~$\delta$ (H$\delta$), rare-earth, and molecular transitions.

\section{Observations} \label{sec:obs}

SUSI is a scanning slit spectropolarimeter that observes in the wavelength domain between 309 and 417\,nm. 
The observations were acquired from 12:18 to 12:55\,UT on 2024 July 16 \citep[\mbox{Sunrise ID:} \mbox{37\_LIMB}, see][]{Solanki2026...sunrise}. The wavelength range is \WavelengthRange{}\,nm. The spectral point spread function has a measured full width at half-maximum of 44.0\,m\AA{}. We obtained this value by matching the Fourier transform spectrometer (FTS) atlas \citep{Neckel1999SoPh..FTS, Hoelken24} to the flat field of the observations. The spectrum is sampled with 13.9\,m\AA{} per pixel. For further details, see section 4.3 of \citet{Feller2025SoPh..SUSI}. For this study, we used only Stokes $I$ spectra.

The slit-scan mechanism of SUSI operates in a continuous mode, allowing the pixel size perpendicular to the slit to be chosen during the data reduction process. To obtain approximately square pixels, data were binned along the scan direction at 3.3\,s intervals. After binning, the dataset comprises 723 scan positions, yielding a field of view of 19\farcs2$\times$44\farcs2 along the scan and slit directions, respectively, with a spatial sampling of 0\farcs0297 per pixel. The scan started near the eastern limb inside the solar disk at (-860\arcsec{},360\arcsec{}), with the slit almost parallel to the limb, and proceeded until it was 10\arcsec{} outside the limb. 

A snapshot of SUSI observations is displayed in Figure~\ref{fig:observations} and in the accompanying \href{\animation{}}{online animation}. During the observations, the CWS lost the locking point twice at 12:31 and 12:33\,UT but quickly regained it a minute later at a slightly different position, allowing the observations to continue without stopping the scan. We lost 26 (0\farcs77) and 18 (0\farcs53) slit-scan positions due to two CWS pointing losses. Because the slit was tilted with respect to the limb, we could easily determine its position before and after each jump. After flattening the limb, the jumps affected different parts of the slit at different distances from the limb. One-third of the slit experienced the jumps far inside the disk, the middle part near the limb, and the final part sat at the limb. To avoid any effects from the jumps near the limb, we only used the third of the slit where the jumps occurred completely inside the disk. This corresponds to 650 out of the 1680 total spectra along the slit.

Despite the almost space-like observing conditions in the stratosphere, the Stokes $I$ spectra are not completely stray light-free. At the end of the scan, when the slit is furthest away from the limb (10\arcsec{}), the spectrum does not show any solar signature of the spectral lines anymore, and contains only residual stray light. The averaged spectral profile recorded at the end of the scan was therefore removed from every profile closer to the limb and on disk. The main contributors to this stray light are readout noise and photon noise, and it amounts to approximately 0.002\% of the average intensity at the limb (see below).

\begin{figure*}
    \centering
\includegraphics[width=.999\linewidth]{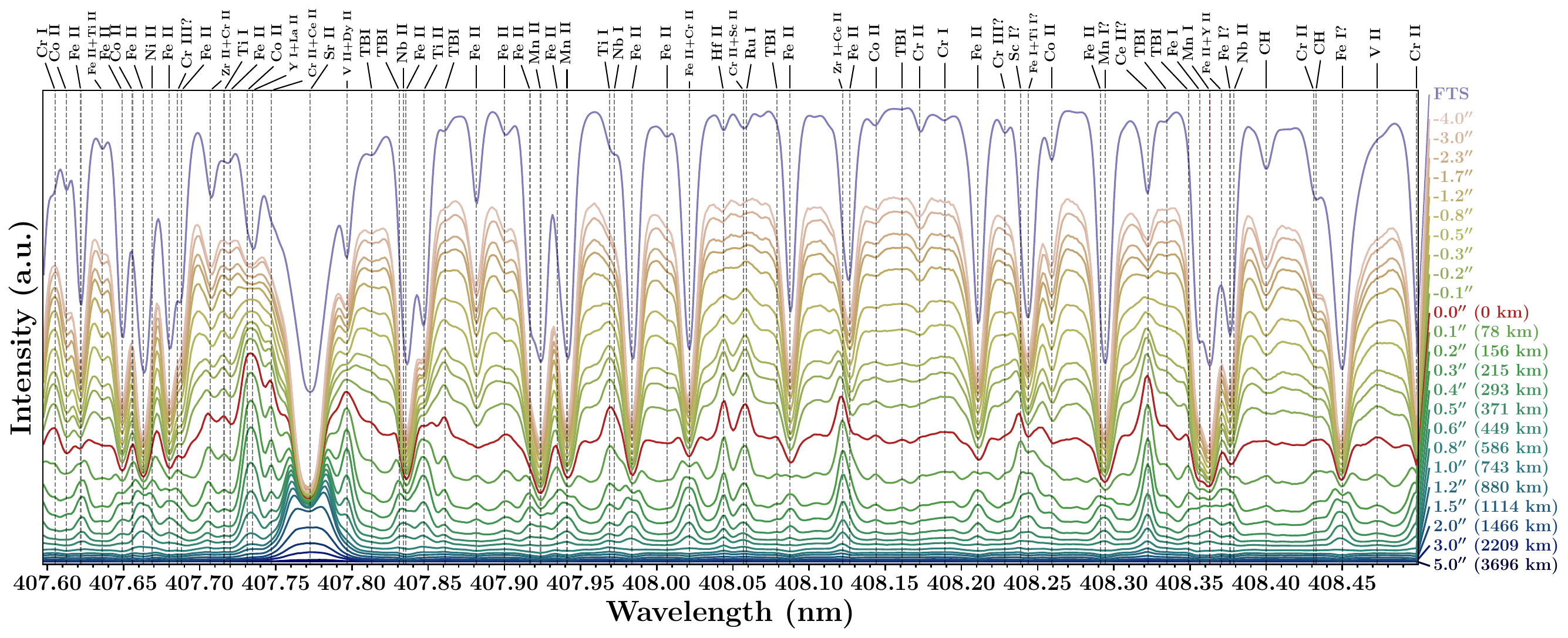}\\[-0.15cm]
\includegraphics[width=.999\linewidth]{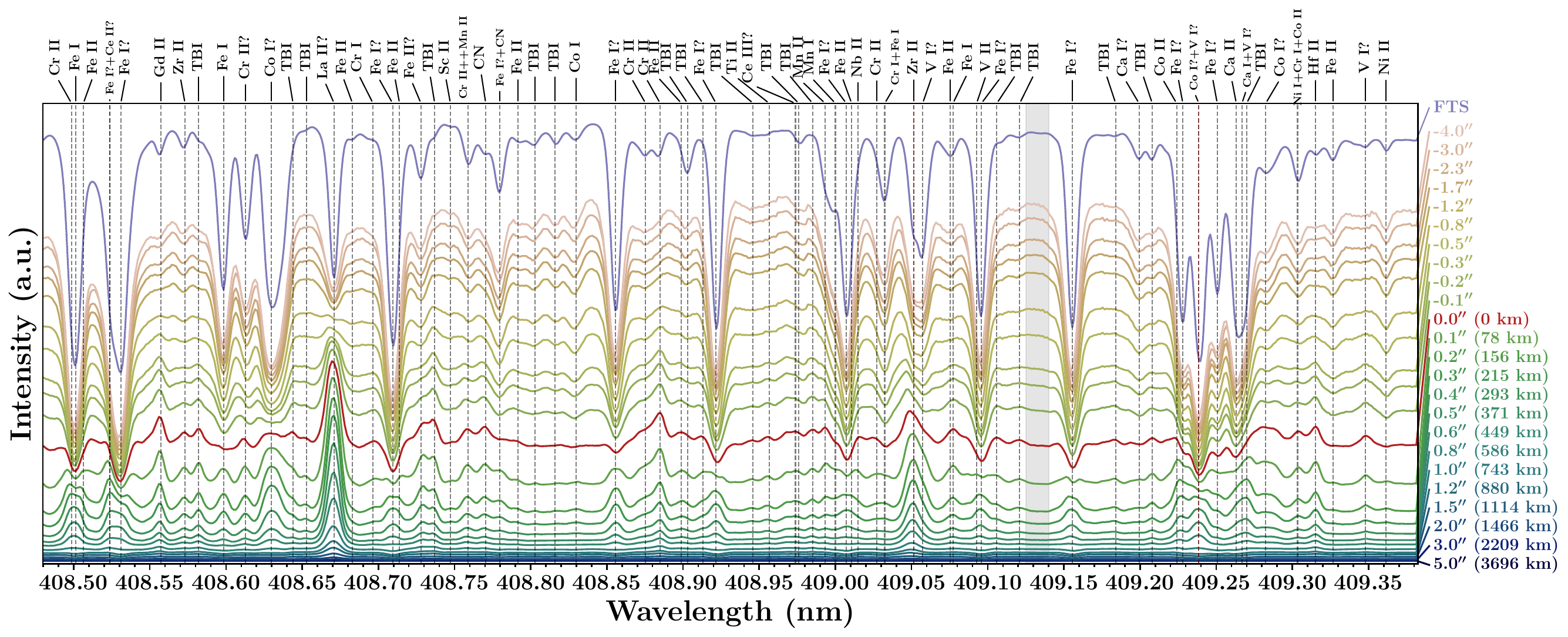}\\[-0.15cm]
\includegraphics[width=.999\linewidth]{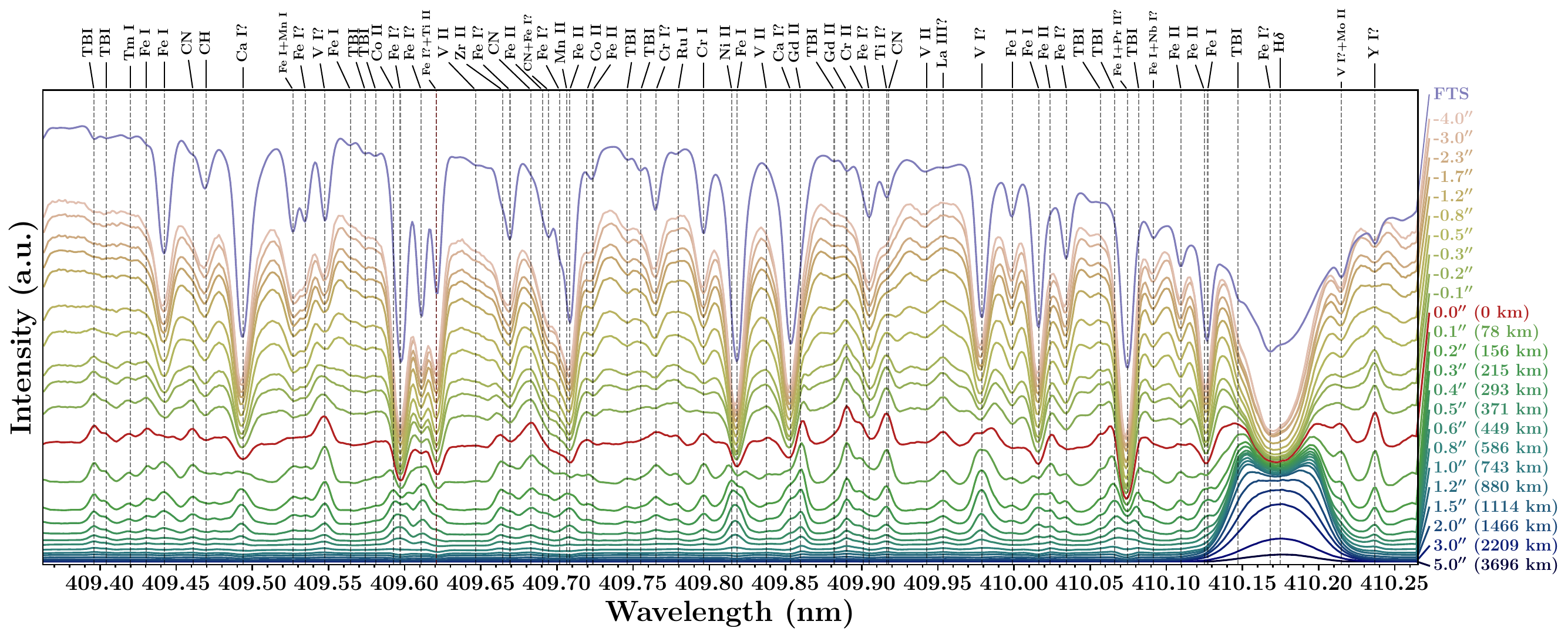}

    \caption{Stokes $I$ spectra covering the wavelengths between 407.6 and 410.2\,nm observed by SUSI. The purple line has been taken from an FTS atlas recorded at disk center \citep{Neckel1999SoPh..FTS}. Colored lines below it represent spectra on-disk (denoted by negative heights), and above the solar limb (positive heights). The heights of the spectra are marked in the same color on the right-hand side in arcseconds and in kilometers starting at the solar limb. The spectrum right at the solar limb is highlighted in red. For better visibility, the full wavelength range is broken up into three smaller ranges.}
    \label{fig:spectra}
\end{figure*}

As seen in Figure~\ref{fig:observations}, the slit was not entirely perpendicular to the limb. We straightened the limb by removing its curvature through cross correlation of limb profiles along the slit direction in the pseudocontinuum image produced by averaging the 409.125--409.14\,nm range (shaded in gray in Figure~\ref{fig:spectra}). This small spectral region was chosen because it is the closest to the continuum,  and because it is the only region that shows no clear spectral lines, neither on-disk nor off-limb. Hereafter, based on the pseudocontinuum image, we determine the inflection point of the radial intensity profile as the solar limb and adopt it as the zero level of height, with positive heights above it. Please note that this zero height is distinct from the spatially averaged optical depth unity level often used at disk center. 

%%%%%%%%%%%%%%%%%%%%%%%%%%%%%%%%%%%%%%%%%%%%%%%
\section{Line identification}
%%%%%%%%%%%%%%%%%%%%%%%%%%%%%%%%%%%%%%%%%%%%%%%
To identify the spectral lines in our observations, we used two solar spectrum atlases that cover spatial locations close to the limb: the second solar spectrum \citep{gandorfer2000,gandorfer2002}, which uses the identifications from \citet{Moore1966sst..SolarSpectrum} and \citet{Wallace1998assp...FTSatlas}, and the chromospheric spectrum taken during a solar eclipse by \citet{Pierce1968ApJS...ChromosphericSpectrum}. In addition, we used the line list curated by
\citet{Riethmueller2014A&A...568A..13R, Riethmueller2019A&A...SUSI}, which is a compilation of the Kurucz atomic line list \citep{kurucz1995atomic}, the Vienna Atomic Line Database \citep[VALD;][]{Piskunov1995A&AS...VALD, Kupka2000BaltA...VALD}, the molecular transitions \citep{coxon1980...molecularconstants,Abrams1994ApJS...SpectroscopyOH}, and the line list provided by the National Institute of Standards and Technology \citep[NIST;][]{Kramida2010...NIST}. We also compared the SUSI observations with a recent compendium by \citet{vanHoof2018Galax...AtomicLineList}. Nonetheless, some clearly visible lines remain unidentified. These lines are marked as to-be-identified (TBI) in the following figures.

We identified spectral lines in an average spectrum obtained by averaging 650 spectra along the axis parallel to the flattened limb. The average $1\sigma$ noise of a single spectrum is $\sim$3 in arbitrary units (a.u.). Averaging reduces this noise to $\sim$0.12 a.u. We set a minimum signal threshold of 10 a.u., implying a signal-to-noise ratio of more than 80 for all identified lines. To keep a line in our final list, it had to show a signal in the FTS data (i.e., at disk center), or become visible close to the limb in the SUSI observations. 

We then compared our data with solar atlases and line lists. We primarily used a curated list based on the work of  \citet{Riethmueller2019A&A...SUSI}, which synthesized hundreds of lines, and kept only those that show a measurable signal under solar conditions. We also cross-checked rare-earth, heavy-element, and molecular transitions across resources, ensuring that only those that appear in more than one resource are kept. For lines with available atomic data, we verified that the $\log(gf)$ values were not too small. If a blend pair had the exact same wavelength and appeared in several resources, we kept both lines. All wavelength numbers are in air.

Atomic information across different line lists can vary, have large uncertainties, or be unknown. We identified approximately 37 distinct transitions that appear in multiple line lists with different ionization states between the lists. For example, the La (listed as either \textsc{ii} or \textsc{iii}) transition at 408.671\,nm, the Fe (\textsc{ii} or \textsc{iii}) transition at 408.727\,nm, and the V (\textsc{i} or \textsc{ii}) transition at 409.057\,nm. Since the ionization state does not affect our conclusions, we denote these specific transitions using the lowest ionization state found in the lists and mark them with a question mark. We also identified four transitions belonging to doubly ionized species (two \ion{Cr}{3}, one \ion{Ce}{3}, and one \ion{La}{3}), and marked them, too, with a question mark. Such ionization states are expected to be difficult to find in the lower solar atmosphere due to the comparatively low temperatures there, but they are the only ones in the sources we used that we could relate to the wavelengths of the observed spectral lines.

\section{Results} \label{sec:results}

The spectra in Figure~\ref{fig:observations} correspond to three distinct observed locations: on-disk (green, Figure~\ref{fig:observations}c), at the limb (blue, Figure~\ref{fig:observations}d), and off-limb (orange, Figure~\ref{fig:observations}e). The on-disk spectrum shows mostly absorption spectral lines, with some weak absorption lines beginning to go into emission when approaching the limb. Most of these go into emission already inside the limb (Figure~\ref{fig:observations}b and c). The complex profile of the \ion{Sr}{2} resonance line at 407.77\,nm shows the wings in emission while the self-absorption core is at the pseudocontinuum level (panel \ref{fig:observations}d). Additionally, nearby blends in emission that appear stronger than the \ion{Sr}{2} wings are present. In the blue wing, these blends include \ion{Y}{1}$+$\ion{La}{2} (407.7347\,nm) and \ion{Cr}{2}$+$\ion{Ce}{2} (407.747\,nm), while \ion{V}{2}$+$\ion{Dy}{2} (407.7969\,nm) appear in the red wing.

Figure~\ref{fig:spectra} shows the spectra averaged spatially in the direction parallel to the straightened limb. Because the slit was not exactly parallel to the limb, the averaging was done over multiple slit positions to ensure that the distance to the limb was the same for all spectra being averaged. The plotted spectra start on disk (upper spectra, coloured brown) and transition to off-limb profiles (lower spectra, coloured green). The two broad lines on-disk are H$\delta$ at 410.17\,nm and \ion{Sr}{2} at 407.77\,nm.  During the transition from on-disk to off-limb, these lines show the typical self-absorption profiles of optically thick emission lines up to $\sim$1\,Mm. Above that, they turn into single-peak emission profiles as they become increasingly optically thin, phenomenologically similar to the behavior of the \ion{Ca}{2}~K line, studied in \citet{Milic2026sunrise}.

\begin{figure}[htbp]
    \centering
    
    \includegraphics[width=0.49\textwidth]{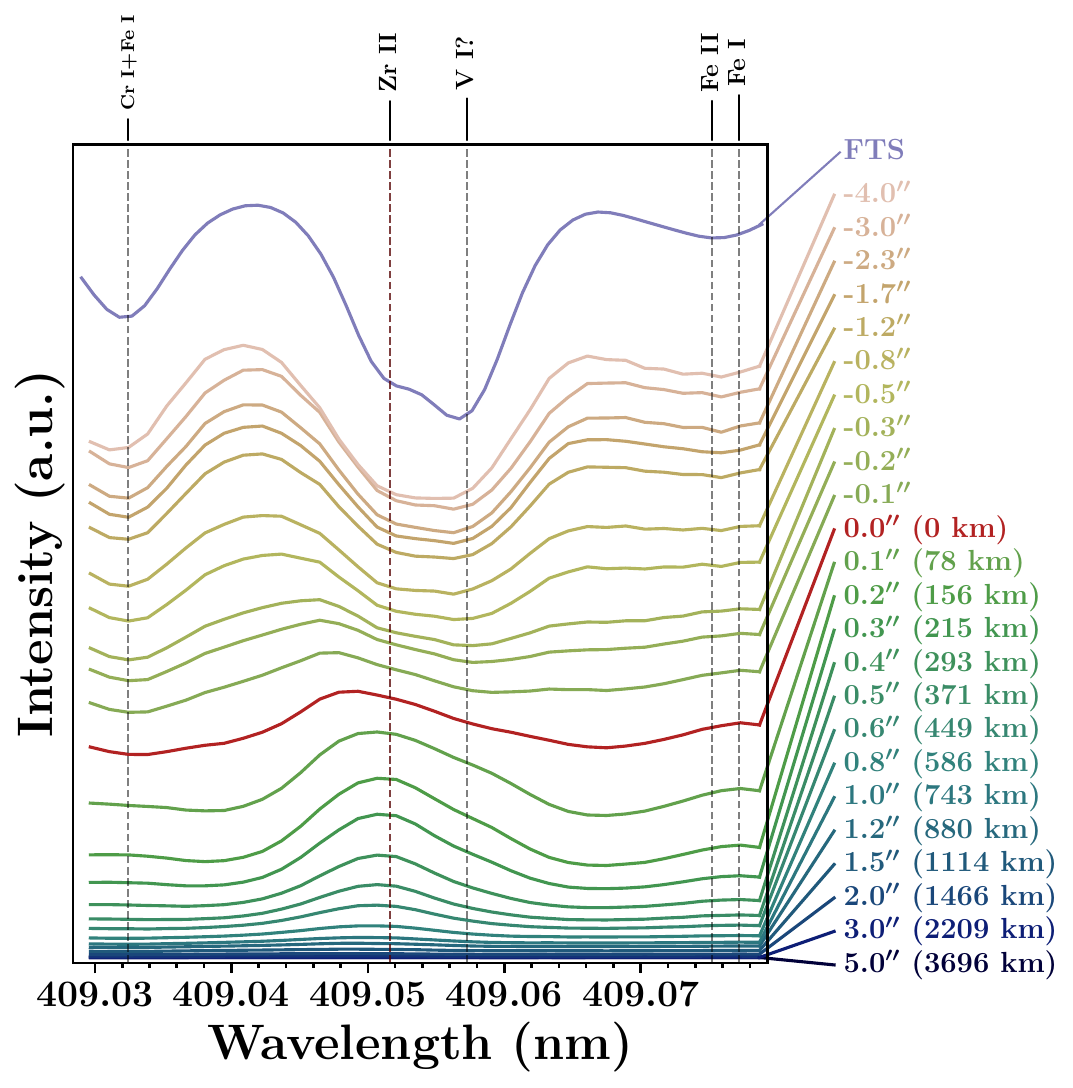}

    \caption{Spectra at different limb distances centered on a blend of two lines, where the \ion{Zr}{2} line, which is weaker at disk center (see the FTS atlas in purple) becomes stronger than the main \ion{V}{1} transition around the limb. Same layout as Figure~\ref{fig:spectra}.}
    \label{fig:blends}
    
\end{figure}

We identified \TotalLines{} distinct spectral lines in total (combining lines identified on-disk and off-limb), belonging to \TotalElements{} different elements or molecular species, spanning an atomic number range from 1 (hydrogen) to  \MaxAtomicNumber{} (hafnium). Of these spectral lines, \TotalLinesOndiskOnly{} lines were observed only on the solar disk, \TotalLinesOffLimbOnly{} lines were observed only off-limb, and \TotalLinesOnBoth{} lines were observed in both locations. Off-limb-only lines (which we interpret as being too weak to be clearly identified on-disk) mostly include ionized transitions (e.g., \ion{Ti}{2}, \ion{V}{2}). There are \UnidentifiedLinesCount{} unidentified lines, which account for \PctUnidentifiedLines{} of the total. These \UnidentifiedLinesCount{} belong to the three groups introduced above. 

Among the total of \TotalLines{} lines, \PctIonizedLines{}\% are ionized, and \PctNeutralLines{}\% are neutral. The majority of the lines belong to \ion{Fe}{1} and \ion{Fe}{2}, both on-disk and off-limb. Twenty lines were identified as the following rare-earth elements: three \ion{Sc}{1} \ion{and}{2}, four \ion{Y}{1} \ion{and}{2}, three \ion{La}{2} \ion{and}{3}, five \ion{Ce}{2} \ion{and}{3}, one \ion{Pr}{2}, three \ion{Gd}{2}, one \ion{Dy}{2}, and one \ion{Tm}{1}, along with two lines for Hf and five for Zr.  Three molecular transitions belonging to the CH molecule and six transitions belonging to the CN molecule were also identified.

The relative strength of the lines appears different on the disk and off-limb. Several examples appear in the observed spectral range. For instance, the \ion{La}{2} line at 408.670\,nm displays strong emission, whereas the nearly equally strong  (at disk center) \ion{Fe}{1} line at 408.598\,nm shows significantly less emission. Additionally, the two \ion{Cr}{2} lines at 408.875\,nm and 408.885\,nm show similar intensities at the disk center, but the line at 408.885\,nm displays strong emission outside the solar disk, whereas the other does not. This effect is also observed in blends. An example is displayed in Figure~\ref {fig:blends}. Blends composed of a line that is stronger and one that is weaker on-disk often show the opposite relative line strengths when in emission off-limb, i.e., the weaker absorption line becomes stronger at some distance above the limb. The distance at which the weaker absorption line becomes stronger varies for each blend. We found this pattern in \weaklines{} blends.

\begin{figure*}[htbp]
    \centering
    \includegraphics[width=1\textwidth]{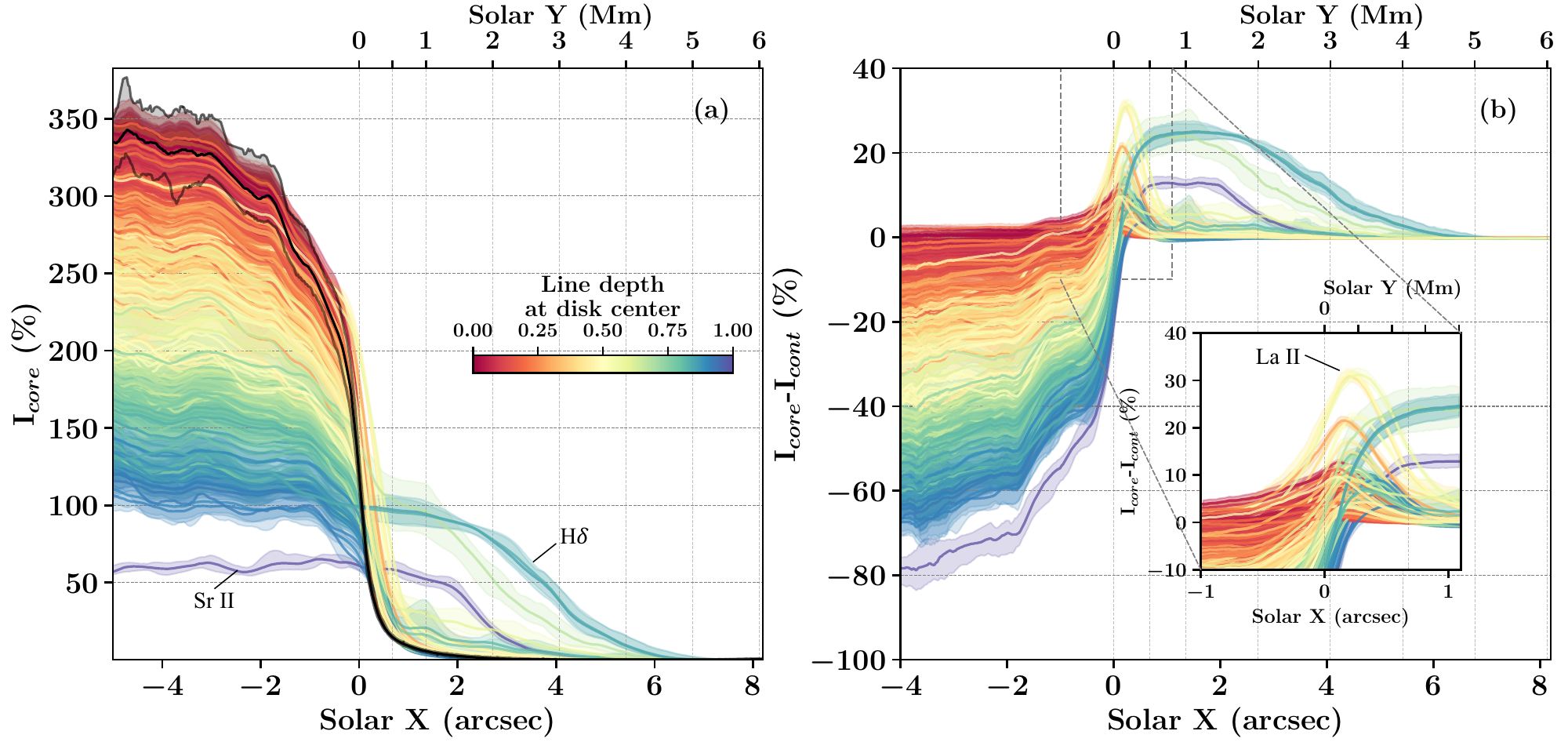}
    \caption{Panel (a) shows the median line-core intensity as a function of limb distance for all lines within the studied wavelength range. The shaded area marks the 5th and 95th percentiles of the line core intensity for a given height after the limb was straightened. The black line and its corresponding gray band are the limb profile obtained from the pseudocontinuum at 409.13\,nm (see shaded wavelength range in Figure~\ref{fig:spectra}). Profiles are color coded by the line depth normalized to the pseudocontinuum taken from the FTS spectrum at disk center, where one is given for the strongest line, and zero represents the pseudocontinuum level (see inset color scale). (b) The same as in panel (a), but after subtracting the pseudocontinuum profile (i.e. the black line in panel (a)). The inset shows a close-up near the limb.}
    \label{fig:lineintensity}
\end{figure*}

\begin{figure}[htbp] 
    % First row
\includegraphics[width=\linewidth]{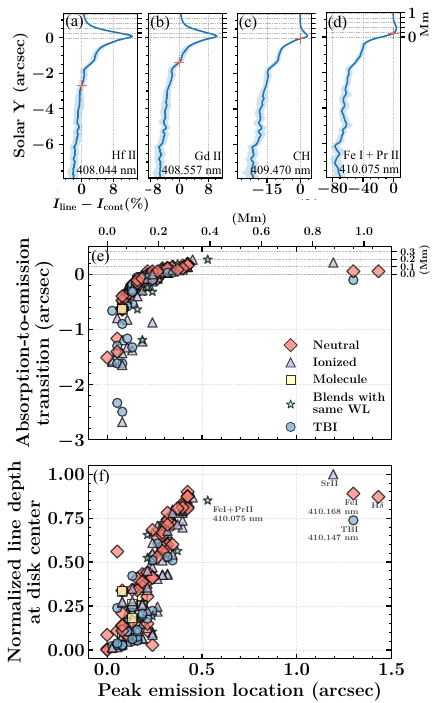}
        \caption{
        (a)-(d) Four examples of the line-core ($I_{\rm core}$) intensity profiles after subtracting the pseudocontinuum limb profile ($I_{\rm cont}$) in the direction tangential to the flattened limb as a function of limb distance. Dark blue marks the median, and the shaded area shows the 5th and 95th percentiles. The red cross marks where $I_{\rm core}$ equals $I_{\rm cont}$, indicating the limb distance where the line core starts turning into emission. Panels (e) and (f) display scatter plots of the limb distance where the different lines start going into emission (panel e) and the normalized line depth at disk center (panel f) as a function of the limb distance where the line core emission peaks. The symbols indicate whether the line is neutral, ionized, a molecule, a blend with the same wavelength but different ionization, e.g., \ion{Fe}{1}$+$\ion{Pr}{2}, or TBI. } 
    \label{fig:weaklines}
\end{figure}

In Figure~\ref{fig:lineintensity}a, we display the line-core intensity as a function of the distance to the solar limb for each identified spectral line. We have color coded each line by the line depth in the FTS spectrum taken at disk center, where red denotes weak absorption lines whose rest intensities lie close to the pseudocontinuum level, and blue denotes the deepest line in this spectral region at disk center, which is the resonance \ion{Sr}{2} line at 407.77\,nm. In Figure~\ref{fig:lineintensity}b, we plot the line core intensity after subtracting the pseudocontinuum intensity at the same spatial location. Clearly, most of the lines undergo a transition from absorption to emission. The limb distance at which this transition occurs spans about 1\arcsec{} around the extreme limb. A total of \totalLinesEmissionOndisk{} (\PcttotalLinesEmissionOndisk{}) of the lines are observed in emission already inside the limb. Among these lines, \totalLinesEmissionOndiskTBI{} are unidentified.
Among the remaining \totalLinesEmissionOndiskIdentified{}, \PtctotalLinesEmissionOndiskIdentifiedionized{} are ionized. Also, eight out of nine molecular transitions are detected in emission on the disk. The remaining molecular transition is blended. The inset in Figure~\ref{fig:lineintensity}b shows that the emission of the majority of the lines peaks at a distance of less than 0.5 Mm above the limb.  The median line-core intensity of \hdelta{} reaches heights exceeding 5 Mm above the limb, indicating that it forms in extended chromospheric structures \citep[e.g.,][]{Beck2016SoPh...Limb}.

The top row of Figure\,\ref{fig:weaklines} shows the line-core intensity profiles as a function of limb distance for several interesting spectral lines from the observed spectral region. \ion{Hf}{2} (Figure~\ref{fig:weaklines}a) and \ion{Gd}{2} (Figure~\ref{fig:weaklines}b) lines are barely visible on the disk but are observed in emission up to 800\,km above the limb. The CH molecular line is visible in emission up to 400\,km above the limb (Figure~\ref{fig:weaklines}c). \ion{Hf}{2} and \ion{Gd}{2} turn into emission inside the solar disk, while the CH line turns into emission at the limb. Finally, the \ion{Fe}{1}$+$\ion{Pr}{2} blend (panel \ref{fig:weaklines}d) appears as an emission line with a self-absorption core, which remains below the pseudocontinuum level up to 200\,km. At these heights, most other lines are fully in emission (see Figure\,\ref{fig:observations}d). Recall that the line core may not always be the best way to identify when the line is in emission, since its wings go into emission first, especially in stronger lines.

Next, we show, for all the lines, the distance to the limb at which they first go into emission versus the distance where the emission reaches its maximum (Figure~\ref{fig:weaklines}e). The photospheric lines go into emission closer to the limb, often even on the disk, and their emission peaks closer to the limb.  In addition, Figure~\ref{fig:weaklines}f shows that the location of the peak emission correlates strongly with the line depth at disk center. That is, weaker absorption lines exhibit their emission peaks closer to the limb, and this relation is seemingly linear. This behavior does not depend on whether the transition is of a neutral or ionized species, or of a molecule. It also does not depend on whether the transitions are of rare-earth or heavy elements (not shown). Not surprisingly, the strong \ion{Sr}{2} resonance line and H$\delta$ do not follow the trend. The other two lines appearing on the top-right of Figures~\ref{fig:weaklines}e and \ref{fig:weaklines}f, \ion{Fe}{1} (410.168\,nm) and TBI (410.147\,nm), are heavily blended near the core of H$\delta$, which likely affects their position in Figure~\ref{fig:weaklines}.

\section{Discussion and conclusions}\label{sec:diskussion}

We report observations of extreme-limb and off-limb spectra in the spectral range 407.6--410.25\,nm. We find \TotalLinesOffdiskOnly{} emission lines just outside the solar limb, many of them remaining clearly visible up to a few hundred kilometers above the limb. Surprisingly, \totalLinesEmissionOndisk{} of these lines, a significant fraction of all off-limb emission lines, begin to appear in emission already on the disk. The identification of the emission lines reveals that the majority are due to \ion{Fe}{1} or \ion{Fe}{2} transitions, about one-fifth of the total  belong to rare-earth elements or other heavy elements, while a few are molecular transitions. Note that line classification is challenging and occasionally results in misclassifications, particularly for exotic transitions \citep[e.g.,][]{Kastner1995ApJS..LineIdentification}. Yet only these lines had wavelengths that matched the observed spectral lines in the sources (line lists) used.

The seminal ground-based spectroscopic work on the flash spectrum by \citet{Pierce1968ApJS...ChromosphericSpectrum} identified ``approximately 11,500 chromospheric lines" within the 340--1000\,nm range. In particular, in the spectral window from 370 to 900\,nm, more than 3400 appear in emission \citep{Pierce1965PASP...FlashSpectrum}. In the \WavelengthRange{}\,nm spectral range \citeauthor{Pierce1968ApJS...ChromosphericSpectrum} found $\sim$70 chromospheric lines in emission. The forest of \TotalLinesOffdiskOnly{} mostly photospheric emission lines found off-limb represents a 2.5-fold improvement, resulting from advances in instrumentation and favorable stratospheric observing conditions.

Most of the off-limb emission lines found by \citet{Pierce1968ApJS...ChromosphericSpectrum} have clear on-disk absorption counterparts. Clear examples showing the transition from absorption to emission off-limb are the photospheric iron pair at 630\,nm, as demonstrated observationally by \citet{Lites2010ApJ...Limb} and modeled by \citet{Shelyag2015ApJ...Limb}. Using extreme-limb observations of the north solar pole with the slit tangential to the limb,  \citet{Livingston1974SoPh...ExtremeLimb} reported two lines,  \ion{Ni}{1} (585.776\,nm) and \ion{La}{2} (586.375\,nm), without a clear counterpart on-disk, within the observed 40\,\AA{} spectral range. They also reported one line, \ion{Ca}{1} (586.757\,nm), with no emission off-limb, as well as a number of spectral lines observed that remain in absorption off-limb. Our work found a total of \TotalLinesOffLimbOnly{} lines that appear only off-limb without a detectable on-disk counterpart,  and a total of \TotalLinesOndiskOnly{} that show no emission off-limb, out of \TotalLines{} lines identified in the observed spectral range.

Extreme solar-limb observations have revealed intriguing, very weak emission lines in the broad wings of the \ion{Ca}{2} H and K lines \citep{Livingston2011SoPh..WeakEmissionLine}. While these features are barely perceptible and often hidden in the strong wings of the \ion{Ca}{2}~H and K lines near the center of the solar disk, their intensity increases dramatically as the line of sight approaches the extreme limb. We have uncovered 122 more lines that go into emission already on disk, albeit close to the limb. This includes many lines that are relatively unblended, i.e., whose formation heights are not boosted by blending lines, as was the case for the lines in the \ion{Ca}{2} H and K wings observed by  \citet{Livingston2011SoPh..WeakEmissionLine}.

One application of this behavior is to detect hidden blends common in the UV. Examples include the blends around 409.818 and 409.853\,nm. These lines appear relatively narrow and reasonably symmetric in the FTS spectrum at disk center, but show strongly asymmetric off-limb emission profiles due to the presence of hidden blends.

Our analysis also revealed that (1) weaker absorption lines go into emission and reach peak emission (i.e., emission relative to the continuum intensity at the same limb distance) closer to the disk center, and (2) lines that go into emission further inward on the disk also peak closer to the limb. The latter exhibits a seemingly linear relationship between line depth at disk center for photospheric lines, which is not surprising, as weaker lines generally form deeper. These relationships are independent of whether the species is neutral or ionized, or of the element, etc., and do not apply to the strong \ion{Sr}{2} resonance line and H$\delta$.

The spectral region containing the \hdelta{} line has rarely been studied in the literature. Only a few prominence studies have used this line \citep{Wiehr2019ApJ...prominences,Wiehr2025...ionizationmemoryplasmaemiters}, but the authors mainly analyzed the resonance \ion{Sr}{2} line at 407.77\,nm to study partially ionized plasma in prominences. Low-stray-light observations by \sunrise{} enable direct measurement of faint emission, from the low photosphere up to about 5\,Mm (in this particular spatial scan), highlighting its potential as an unexplored proxy for investigating the chromosphere in off-limb observations.

The present work has revealed the richness and the complexity of the off-limb spectrum in the violet spectral range, with spectral lines displaying partly quite different behaviors. This exploratory work complements recent off-limb studies on the infrared \ion{Ca}{2} and Paschen 15-17 lines using SCIP observations \citep{Naito2026sunrise, Haocheng2026ApJL}. Gaining more information and understanding will require a more detailed analysis and supporting model calculations such as those presented by \citet{Shelyag2015ApJ...Limb} and \citet{Milic2026sunrise}.

\begin{acknowledgments}
\sunrise{} is supported by funding from the Max-Planck-Förderstiftung (Max Planck Foundation), NASA under Grant \#80NSSC18K0934 and \#80NSSC24M0024 ("Heliophysics Low Cost Access to Space" program), and the ISAS/JAXA Small Mission-of-Opportunity program and JSPS KAKENHI Grant Numbers JP18H05234 and JP23K25916. This research has received financial support from the European Union's Horizon 2020 research and innovation program under grant agreement No. 824135 (SOLARNET) and No. 101097844 (WINSUN) from the European Research Council (ERC). It has also been funded by the Deutsches Zentrum für Luft- und Raumfahrt e.V. (DLR, grant no. 50 OO 1608). The Spanish contributions have been funded by the Spanish MCIN/AEI/10.13039/501100011033 under projects RTI2018-096886-B-C5, PID2021-125325OB-C5, and PID2024-156066OB-C5, and from "Center of Excellence Severo Ochoa" awards to IAA-CSIC (SEV-2017-0709, CEX2021-001131-S), all co-funded by "ERDF A way of making Europe". F. A. Iglesias is a member of the “Carrera del Investigador Científico” of CONICET and supported by MPG through the Max Planck Partner Group between MPS and the University of Mendoza, Argentina. 
\end{acknowledgments}

\bibliographystyle{aa}

\bibliography{references,sebas_references}

\end{document}